\documentclass[paper]{revtex4}

\usepackage{graphicx}
\usepackage{epsfig}
\usepackage{fancyhdr}
\usepackage{empheq}
\usepackage{mathbbol}
\usepackage{wasysym}
\usepackage{pstricks}
\usepackage{color}
\usepackage{bbm}
\usepackage[utf8x]{inputenc}
\usepackage{amssymb,amsmath}
\usepackage{epstopdf}
\usepackage{float}
\usepackage{color}
\usepackage{subfig}
\graphicspath{{./immagini/}}

\begin{document}

\title{The $\Delta$-Statistics of Uncoventional Quarkonium-like Resonances}


\author{ENM Cirillo$^{\dag}$, M Mori$^{*}$ and AD Polosa$^*$}
\affiliation{$^\dag$Dipartimento di Scienze di Base e Applicate per l'Ingegneria, Sapienza Universit\`a di Roma, I-00161 Roma, Italy\\
$^*$Dipartimento di Fisica, Sapienza Universit\`a di Roma, Piazzale A Moro 2, Roma, I-00185, Italy}
\begin{abstract}
In this note we address the problem of unconventional charmonium-like levels from the standpoint of level spacing theory. 
The level distribution of the newly discovered vector resonances is compared  to that of standard charmonia analyzing 
their spectral rigidities. It is found that the unconventional charmonium-like states are significantly more compatible with 
the hypothesis of being levels from a Gaussian Orthogonal Ensamble of Random Matrices  than 
the standard ones, which in turn seem more likely to be Poisson distributed. We  discuss the consequences 
of this result and draw some hints for future investigations.
\end{abstract}
\maketitle

{\bf \emph{Introduction}}. 
With the very recent observation of the charged resonances $Z(10610)$ and $Z(10650)$ by the Belle collaboration~\cite{belle}, the family of 
unconventional quarkonium-like states has further grown. Since the discovery of the $X(3872)$, now observed also in LHC experiments,
a long list of new narrow resonances has been found. There is a vast consensus that most of them are multiquark structures
although a general picture is still missing. Some of these states occur extremely close and some other far from open-charm or beauty thresholds.
For the close-to-thresholds ones, several authors agree that the appropriate interpretation is in terms of $S-$wave
$D^{(*)}D^{(*)}$ or $B^{(*)}B^{(*)}$ hadron molecules with a very small binding energy (compatible with zero), yet rather stable to be as 
narrow as the observation shows. Also the prompt production of  $X(3872)$ in $p\bar p$ collisions at CDF has been observed  making at least questionable the 
chances of a loosely bound molecule interpretation~\cite{noiben}. On the other hand it has been claimed that final state interactions
mechanisms could be at the core of the surprising stability of such a molecular object~\cite{abrat}.  Similarly the newly discovered states, the $Z(10610)$ and 
$Z(10650)$, have immediately been interpreted as hadron molecules~\cite{volo} for their mass values happen to be exactly at the 
threshold values of $BB^*$ and $B^*B^*$ mesons. 

Hadron molecules are meant to be {\it extended tetraquark objects} (several fermi in size) in which the strong interaction is conveyed by some long range
pion exchange or rescattering mechanism. As opposed to this picture one could theorize the existence of {\it compact tetraquark} structures
which are  just new kind of hadrons with four quarks neutralizing the color within the typical range of strong interactions~\cite{mainoi}.
In principle, compact tetraquarks are not expected to be formed at the mass values of  meson molecules; on the other hand some $Qq\bar Q\bar q$ 
bound state could fluctuate into $(Q\bar q)(\bar Q q) $ or $(Q\bar Q)(\bar q q)$ and the discrete levels of the unknown Hamiltonian binding $Qq\bar Q\bar q$
should, as a result, be  coupled to hadron molecule levels.

In this letter we test the assumption that the known $1^{--}$ resonances located  at the mass values ${\cal E}=\{3943, 4008, 4263, 4360, 4634, 4664\}$~MeV -- 
all of them candidates to be exotic hadrons~\cite{nb} -- represent the discrete levels of some unknown compact tetraquark Hamiltonian along the same lines 
as the standard charmonia at  ${\cal S}=\{3096~(J/\psi), 3686~(\psi(2S)), 3772, 4039, 4153,  4421\}$~MeV are the levels of the $c\bar c$ Hamiltonian with the Cornell potential.
Resonances in ${\cal E}$ are all produced in $e^+e^-$ collisions with initial state radiation.
Most of the levels of the exotic set ${\cal E}$ happen to be away from open charm threshold and  thus represent a good laboratory to explore 
the possibility that we are observing the spectrum of a complicated multiquark Hamiltonian; for earlier attempts of this kind see~\cite{vv}.  

A very much studied conjecture in the field of quantum
chaotic systems~\cite{bohigas84}, states that 
{\it spectra of quantum  Hamiltonian systems whose classical analogs are described by (strongly) chaotic Hamiltonians
show locally the same fluctuation properties as predicted by the so called Gaussian Orthogonal Ensemble~(GOE) for large dimensional Random Matrices}.
A portion of the quantum Hamiltonian spectrum is rescaled to spacing one and the levels so obtained turn out to be distributed as the eigenvalues of the
GOE Random Matrices in the limit of large dimensions. 
In this limit indeed the local properties of Random Matrix eigenvalues are extracted, as the Wigner semicircle appears locally flat. 
As a consequence the probability distribution of the level spacings is expected to follow closely the Wigner law
\begin{equation} \label{wigner distribution}
W (s)= \frac{\pi s}{2} \, \exp \Big( {-\frac{\pi s^2}{4}}\Big) 
\end{equation}
The eigenvalues of GOE matrices following this distribution show the typical {\it level repulsion} features studied at length in the context of nuclear resonances.

Naive formulations of tetraquark semiclassical 4-body Hamiltonians are possible, for example relying on one-gluon-exchange models. 
Most likely all of them express a chaotic classical dynamics. The Hamiltonian describing the $c\bar c$ system is, on the other hand, very close to an integrable one.
Thus it is expected to have the level clustering features of the Poisson spacing distribution $P(s)=\exp(-s)$, or at 
least a discrepant behavior with respect to  the GOE eigenvalues.

Using the tool of the {\it spectral rigidity} also known as the $\Delta-$Statistics developed initially by Dyson and Mehta we study the short sets
${\cal E}$ and ${\cal S}$ in the attempt of confirming or disproving the picture according to which the ${\cal E}$ levels should more markedly
match the expected behavior for the  GOE ones than standard charmonia, ${\cal S}$, do. We surprisingly find  that this is indeed the case although 
our explorative analysis has its natural limit in the very limited amount of data at hand - the method of $\Delta-$Statistics has been systematically  
applied for example in  the discussion of nuclear resonances level spacing where the data sets contain order of hundreds of levels.

Yet we believe that this result is to be interpreted as an interesting suggestion which leads  us to some speculative considerations we are still working on: 
$i)$ exotic hadrons (for example those in the ${\cal E}$ set) are just 
like the $c\bar c$ ones but with an additional light quark $q\bar q$ component; $ii)$ they fall on the levels of some tetraquark Hamiltonian; 
$iii)$ once a discrete tetraquark level happens to be located within the level width of a  molecular level - centered at some meson; 
threshold -  because of the coupling between the two spectra induced by fluctuations like  $Qq\bar Q\bar q\to (Q\bar q)(\bar Q q) \to Qq\bar Q\bar q$,
the molecular level, {\it otherwise very broad}, gets metastable because of a Feshbach-like mechanism. 

{\bf \emph{$\Delta$--Statistics}}. 
The spectral rigidity (SR) is a measure of the deviation of a level set from  uniform spacing: the more regular the set,
the smaller the value of the spectral rigidity.
Consider a set of $N$ levels $\{ E_i\}$  rescaled to unit spacing, namely $E_N -E_1 = N-1 \equiv 2L$, and centered with respect to the origin ($E_1=-L$ and $E_N=L$).
The sample cumulative function is
$$ C(x) \equiv \sum_{E_i>0}\Theta \left( x-E_i\right)-\sum_{E_i<0}\Theta(E_i-x) $$ 
where $\Theta(x)=1$ if $x\ge 0$ and $\Theta(x)=0$ if $x<0$. The spectral rigidity, in its original form due to Dyson and Mehta, is defined as
\begin{equation} \label{Delta_3}
\Delta_3 \equiv \frac{1}{2L} \min_{A,\,B} \int_{-L}^{L} \left( C(s)-As-B\right)^2 \mathrm{d}s 
\end{equation}
following the notations introduced in~\cite{dysonIV}.

The conjecture above  means that a sequence of $N$ experimental levels has to be compared with a sequence of $N$ eigenvalues extracted from an ensemble of  
random matrices with large dimension $D$. Calculations in~\cite{metha} show that in the large $D$ limit the mean of the SR computed with the Poisson distribution 
is linear in the number of  spacings, while that computed with the GOE~\footnote{As well 
the Gaussian Unitary and Symplectic Ensembles } eigenvalue distribution grows only logarithmically: it is therefore possible
to discriminate between Poisson and GOE levels. This discrimination is more effective as the number of
consecutive levels in the studied sequence grows.
In practice, the number of levels available is often too low for $\Delta_3$ to provide a clear discrimination between Poisson and Gaussian
Orthogonal Ensembles. It is therefore useful to consider a different notion of SR in order to reduce the variance. 
Following Bohigas \textit{et al.}~\cite{bohigas82} we set
\begin{equation} \label{Delta tilde}
\Delta_3 (x,y)\equiv \frac{1}{y} \min_{A,\,B} \int_{x}^{x+y} \left( C(s)-As-B\right)^2 \mathrm{d}s 
\end{equation}
where $\Delta_3 (x,y)$ is a generalization of the Dyson-Mehta estimator, recovered as $y=2L$ and $x=-L$. 
We thus define a new random variable $\Lambda (w,y)$ built on averaging
the spectral rigidity of smaller portions of the dataset
\begin{equation} \label{Delta}
\Lambda (w,y) \equiv \frac{1}{2L-y+2w} \int_{-L-w}^{L+w-y}  \Delta_3 (x,y) \mathrm{d}x 
\end{equation}
where $y$ takes continuous values between $0$ and $2L+2w$, which is the number of spacings of the sequence, while $x$ ranges
from $-L-w$ to $L+w-y$. We have thus defined a family of statistical variables depending on $y$ and $w$.  $\Delta_3$ looks
at the whole data set whereas $ \Delta_3(x,y)$ checks	 a smaller number of levels and $\Lambda(w,y)$ is his average on the data set.
The parameter $w$ is introduced in order to minimize possible finite size effects.
The original definition of $\Delta_3$ is recovered in the limit $w=0\,,\;y \rightarrow 2L $.

%The spectral rigidity of the experimental levels is compared with a large number of sampled GOE and Poisson sequences:
%this choice is motivated by the presence of large finite size effects, which are absent in the Ensemble averages.

{\bf \emph{Results}}. The data sets at hand are ${\cal E}=\{3943, 4008, 4263, 4360, 4634, 4664\}$~MeV,
the candidate exotic levels, and the standard charmonia at  ${\cal S}=\{3096~(J/\psi), 3686~(\psi(2S)), 3772, 4039, 4153,  4421\}$~MeV.
${\cal E}$ contains the masses in MeV of the $1^{--}$ states $G(3900)\, , \;
Y(4008) \, , \;  Y(4260)  \, , \;  Y(4360)  \, , \;  X(4630)$ and $Y(4660)$.
${\cal S}$ contains the masses of the standard $1^{--}$ charmonia, from the $J/\psi$ to the $\psi (4415)$.
A third set ${\cal E}^\prime=\{3943, 4008, 4263, 4360, 4661\}$~MeV is composed by the same resonances of the first set
${\cal E}$, where the $Y(4630)$ and the $Y(4660)$ are taken to coincide with the $Y_B(4660)$ state,
as proposed in \cite{charmed_baryonium}. In this study we will neglect the uncertainties on the masses.

In order to choose the parameter $w$ we are introducing, we study the behavior of $\Lambda(w,y)$ on some test series. 
We choose $w=1$, because smaller values are insensitive to variations of the spacings at the extrema of the series, whereas greater 
values are useless as they do not  add  further information.

From now on we will use the notation $\Lambda(1,y)\equiv \Lambda(y)$.
As we deal with 5 and 6 level sequences, we generate a large number of $\Lambda(y)$ samples from GOE and Poisson series. The GOE samples
are obtained by diagonalizing 30 random GOE matrices  $4000 \times 4000$ in size, obtaining 24000 series of 5 levels each and 19980 series of 6 levels each; 
the number of Poisson samples is similar (24000 and 20000). The integral in $\Delta_3(x,y)$ is evaluated analytically,
whereas the $\Lambda$'s are obtained by a midpoint rectangle approximation. We choose to evaluate $\Lambda(y)$ for integers and half-integers, with $y\in[0.5,7]$
($\Lambda(0)$ is identically zero). Note that each experimental  data set has to be compared with 
samples of same cardinality.

\begin{figure}[H]
  \centering
    \includegraphics[width=0.6\textwidth]{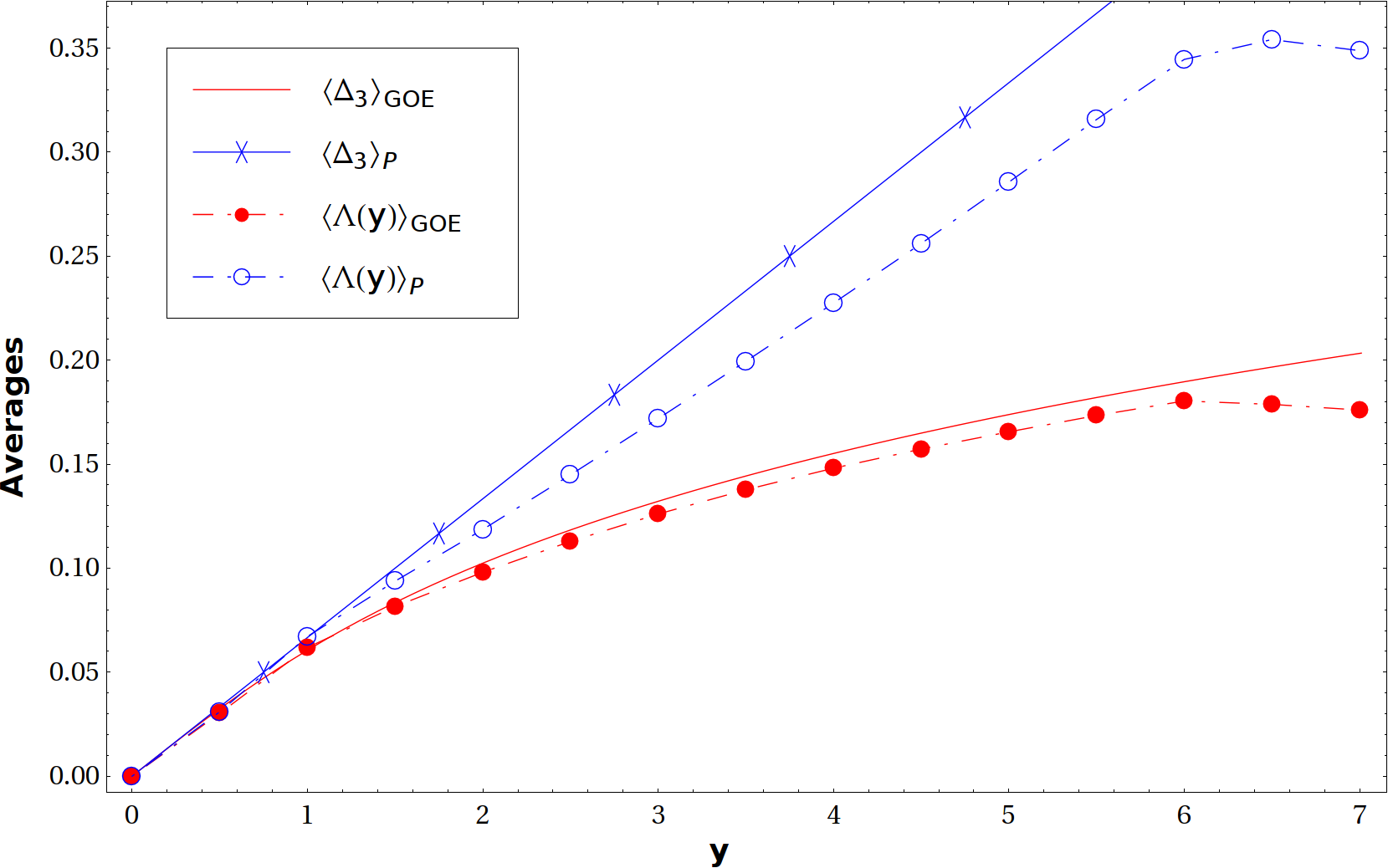}
  \caption{\small{The solid lines represent $\langle \Delta_3\rangle_{_{\rm P,GOE}}$
   for series of $y+1$ levels. The points show the Monte Carlo sampled $\langle \Lambda(y)\rangle_{_{\rm P,GOE}}$.}}
  \label{fig:delta_mean}
\end{figure}

A comparison between the statistical properties of $\Delta_3$ as  a function of the number of spacings and our $\Lambda(y)$ is given in 
Figure~\ref{fig:delta_mean}. There we introduce the ensemble averages
$\langle \cdot \rangle_{_{\rm P, GOE}}$ where P stands for averaging against the Poisson Ensemble and GOE is for the Random Matrix ensemble.
The ensemble average $\langle \Delta_3\rangle_{_{\rm P,GOE}}$ is computed exactly in~\cite{metha}. The  
$\langle \Lambda(y)\rangle_{_{\rm P,GOE}}$ is computed via the Monte Carlo sampling described above.

The $\langle \Lambda(y)\rangle_{_{\rm P,GOE}}$  have respectively  the same linear and logarithmic behavior as $\langle \Delta_3\rangle_{_{\rm P,GOE}}$,
apart from the last points where finite-size effects dominate. A clear discrimination between
GOE and Poisson sets is reached at  $y$ large.
We now study the  level  properties of the experimental series ${\cal E}$ and ${\cal S}$. 
We observe that it is hard to discriminate between Poisson and non-Poisson sets because of the large variance of the
Poisson $\Lambda(y)$ random variable. On the other hand the GOE distributions have a smaller variance, so a more significant discrimination is possible
between GOE and non-GOE sets by looking at high values of $y$.

\begin{figure}
\begin{minipage}[t]{7truecm} % A minipage that covers half the page
\centering
\includegraphics[width=7truecm]{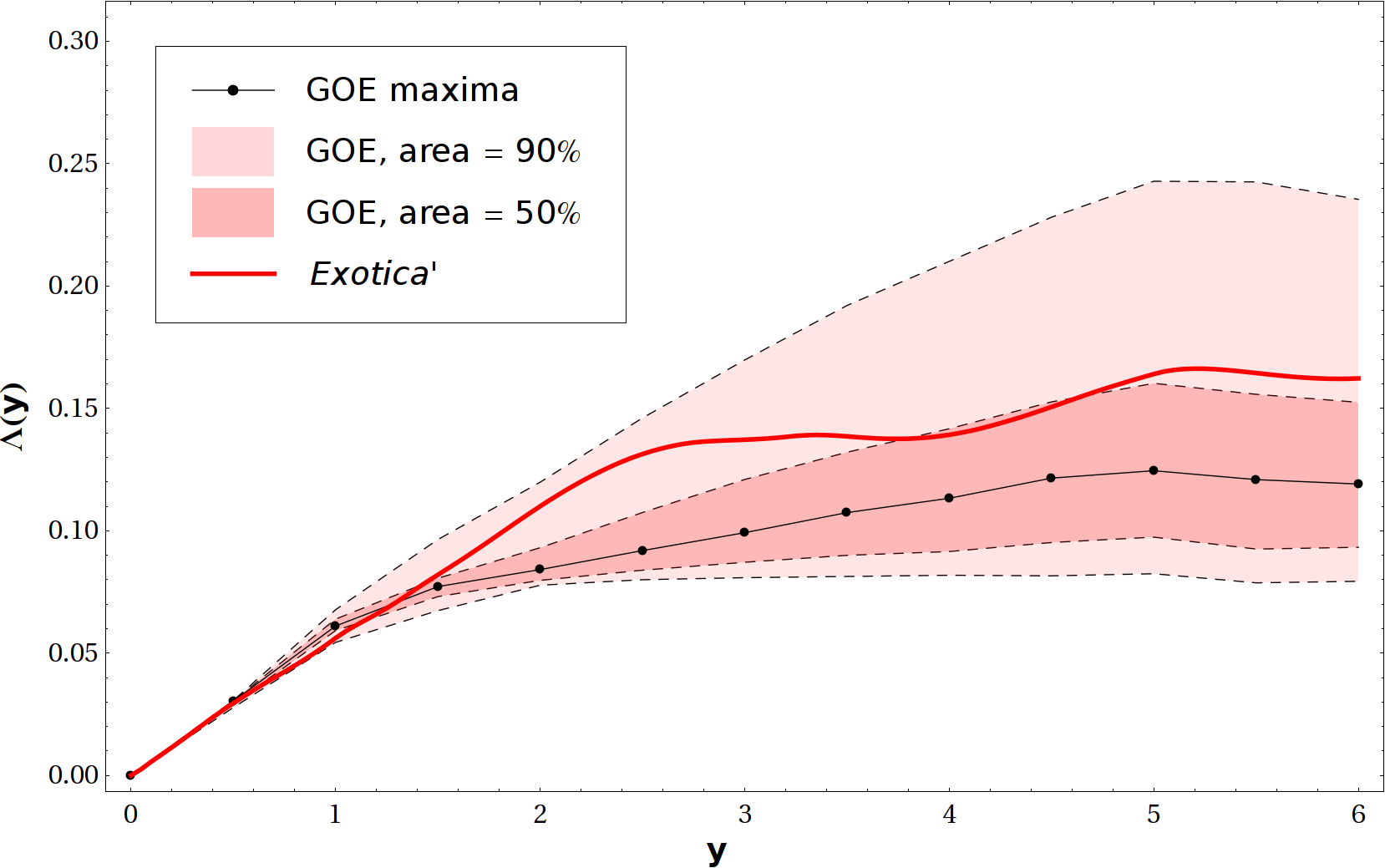}
\end{minipage}
\hspace{1truecm} % To get a little bit of space between the figures
\begin{minipage}[t]{7truecm}
\centering
\includegraphics[width=7truecm]{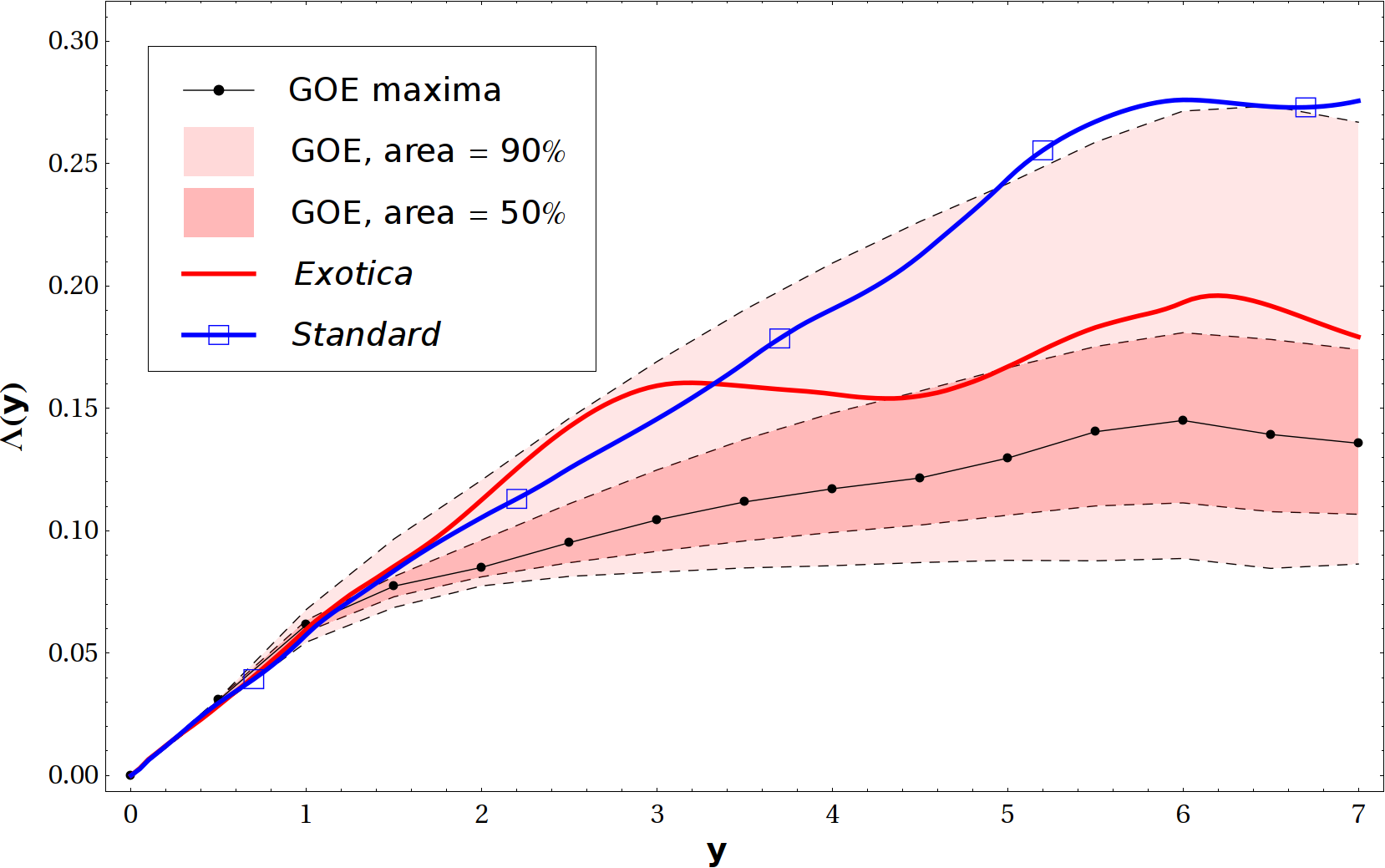}
\end{minipage}
\caption{\small In the two figures the experimental and sampled data for 5-levels (left panel) and 6-levels series are shown.
Solid lines show the experimental $\Lambda(y)$ for the $\mathcal S$, $\mathcal E$ and $\mathcal{E}^\prime$ series.
For any $y$, the black line indicates the point where the maximum of the distribution $f_y$ is attained; the colored areas indicate
the regions that contain 50\% or 90\% of the $\Lambda(y)$ samples. Note that when $\Lambda(y)$ is outside the $90\%$ area, 
the parameter $\alpha_y$ is  smaller than $0.1$.}
  \label{fig:goe_areas}
  \end{figure}

In Fig.~\ref{fig:goe_areas} we show the experimental $\Lambda(y)$ for the ${\cal E}$ and ${\cal S}$ levels compared to the six level GOE averages.
The results obtained for the sets ${\cal E}$ and ${\cal E^\prime}$ are compatible with the hypothesis of GOE distributed levels whereas this turns 
out not to be true for the ${\cal S}$ set.

\begin{figure}[!ht]
  \centering
    \includegraphics[width=0.6\textwidth]{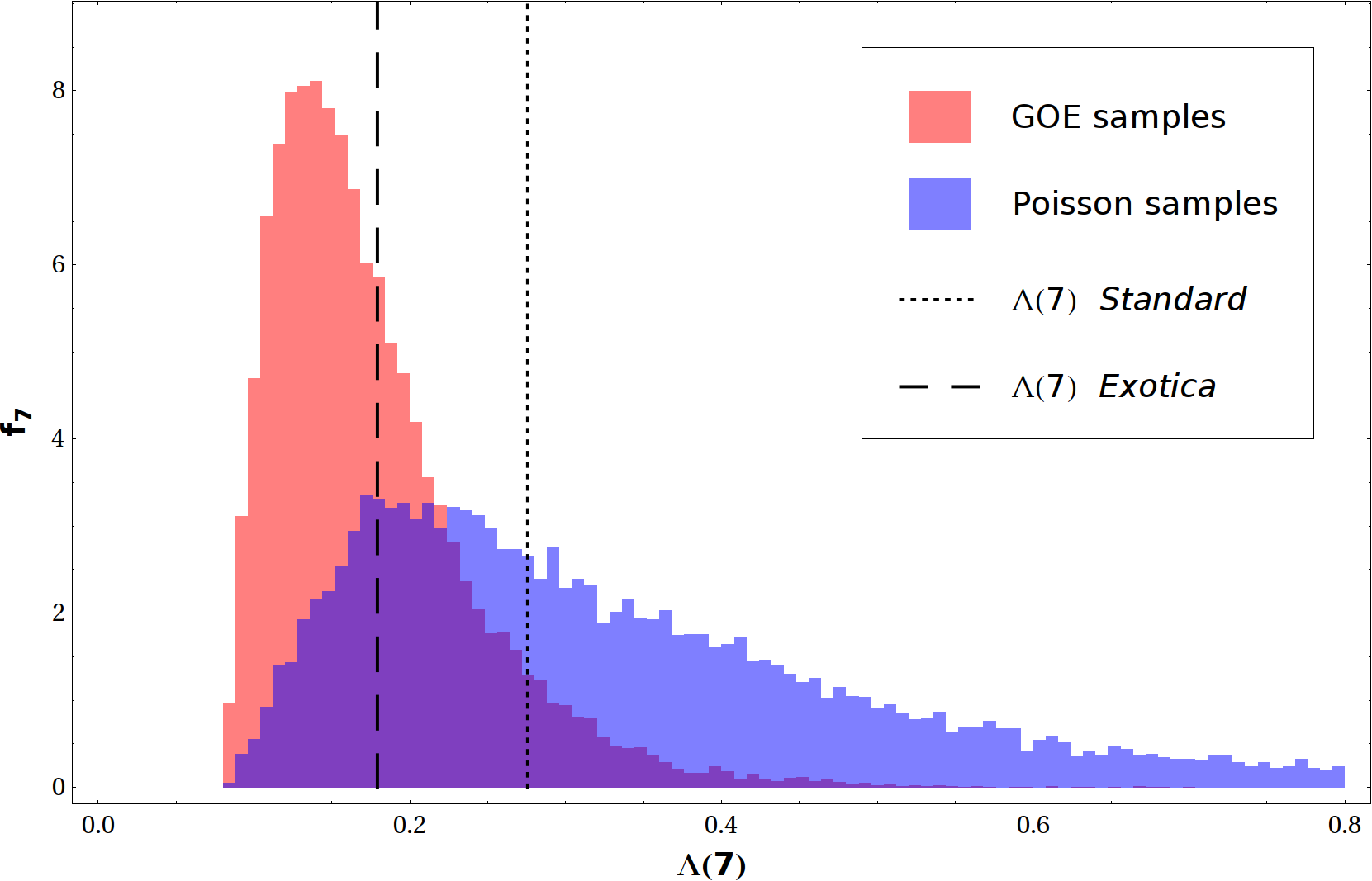}
  \caption{{\small Sampled distributions for $\Lambda(7)$. The histograms are obtained with 19980 GOE samples and 20000 Poisson samples
and rescaled to unit area. The vertical lines indicate the $\Lambda(7)$ value for the \textit{standard} and the \textit{exotica} series.
The bin width is equal  0.008.}}
\label{fig:delta_7}
\end{figure}

Given the random variable $\Lambda(y)$ relative to the GOE ensemble we compute the related distribution function $f_y$ as 
depicted in Fig.~\ref{fig:delta_7}. We thus introduce 
\begin{equation}
 \alpha_y \equiv \int_{\{s : f_y(s)<\bar f_y\}} f_y(s) ds
\end{equation}
where $\bar f_y$ is the value assumed by the distribution $f_y$ in correspondence of the SR  $\Lambda(y)$ associated to the experimental 
data set ${\cal S}$. $\alpha_y$ takes values in the $[0,1]$ interval; in correspondence of small $\alpha_y$ values the null-hypothesis that 
${\cal S}$ is a realization of the GOE ensemble has to be rejected.

We consider the most significant five cases $y=5,5.5,6,6.5,7$ and compute the corresponding $\alpha_y$ by constructing a 
binned distribution $f_y$. By averaging over different binning choices we obtain the data reported in Table.~1 where for $y\ge5$ the fraction $\alpha_y$ is smaller than 0.1. 
\begin{table}[!h]{
  \centering
  \renewcommand{\arraystretch}{1.2}
    \begin{tabular}{|c|c|c|c|c|c|}
    \hline
      $y$ &   $5$ & $5.5$ & $6$ & $6.5$ & $7$ \\
    \hline
    $\alpha_y$ & $0.097 $ & $0.084$ & $0.095$ & $0.096$ & $0.089$\\
    \hline
    \end{tabular}
\caption{ \small{$\alpha_y$ is the fraction of sampled $\Lambda(y)$ whose probability is less than the value $\bar f$ introduced in the text.}}
  \label{alpha_table}}
\end{table}

Computing the analogous quantity for the data sets ${\cal E}$ and
${\cal E^\prime}$ we get values larger than 0.4 for $y\ge 5$.

%This can be partially explained by the strong correlation between close values of $y$. 
It is then reasonable to reject the hypothesis that ${\cal S}$ levels are extracted from the Gaussian Orthogonal Ensemble.
Note that, because of the large variance of the Poisson samples, it is practically impossible to reject
the hypothesis that a series of few levels ($\lesssim 20$) belongs to the Poisson Ensemble, but, as observed above, it is easier to state that they 
are not of the GOE type.

As an example in Fig.~\ref{fig:delta_7} the sampled distributions for $\Lambda(7)$ are shown and the
experimental values of the ${\cal S}$ and ${\cal E}$ sequences.

{\bf \emph{Conclusions}}. 
Comparing the spectral rigidity of standard $S$-wave charmonia with that of the unconventional 
vector resonances recently discovered, we observe that the latter states are more significantly compatible with the hypothesis of being
the levels of some multiquark Hamiltonian (whose classical analog exhibits chaotic dynamics and therefore having quantum levels distributed with the Wigner law) 
than the former, which, on the other hand, could 
be thought as the levels of some classically integrable one - as the simplest version of the Cornell potential Hamiltonian is.
The limit of this analysis is in the small amount of experimental data available both in the standard and unconventional sectors. 
We have introduced a slight modification of the spectral rigidity estimators used in the literature to improve as much as possible
the quality of our analysis with a small number of levels. 

Molecular Hamiltonians, besides the fact that describe two-body systems, could as well be regulated by complicated potentials with Wigner 
distributed quantum level spacings. Yet there are states in the ${\cal E}$
sequence of unconventional $1^{--}$ resonances which do not match molecular thresholds whereas the most accredited hadron molecule model
has $S$-wave  molecules  almost exactly at threshold. There are no clear hints on the form of  these potentials neither and the main  problem of 
the spectroscopy of the new $X,Y,Z$ resonances remains that of finding a unified description that accounts for both   on and off-threshold particles. 

{\bf \emph{Acknowledgements}}. We wish to thank A. Vulpiani for informative discussions.

\end{document}